\documentstyle[11pt,epsfig]{article}

\begin{document}

\title{\bf \Large A dusty pinwheel nebula around \\
                  the massive star WR\,104} 

\author{\bf Peter G. Tuthill, John D. Monnier \& William C. Danchi \\
            Space Sciences Laboratory, University of California,\\
            Berkeley, CA, 94720-7450}
\date{}

\maketitle

\bf
Wolf-Rayet (WR) stars are luminous massive blue stars thought
to be immediate precursors to the supernova terminating their brief lives.
The existence of dust shells around such stars has been enigmatic
since their discovery some 30 years ago; the intense radiation field from
the star should be inimical to dust survival \cite{WHT_AA87}.
Although dust-creation models, including those involving interacting 
stellar winds from a companion star \cite{Usv_MN91}, have been put forward,
high-resolution observations are required to understand this phenomena.
Here we present resolved images of the dust outflow around Wolf-Rayet
WR\,104, obtained with novel imaging techniques, revealing detail on scales 
corresponding to about 40\,AU at the star.
Our maps show that the dust forms a spatially confined stream
following precisely a linear (or Archimedian) spiral trajectory.
Images taken at two separate epochs show a clear rotation with a period 
of $220 \pm 30$\,days.
Taken together, these findings prove that a binary star is responsible 
for the creation of the circumstellar dust, while the spiral plume makes 
WR\,104 the prototype of a new class of circumstellar nebulae unique to 
interacting wind systems.
\rm

\begin{figure}
\begin{center}
\centerline{{\epsfxsize=\columnwidth{\epsfbox{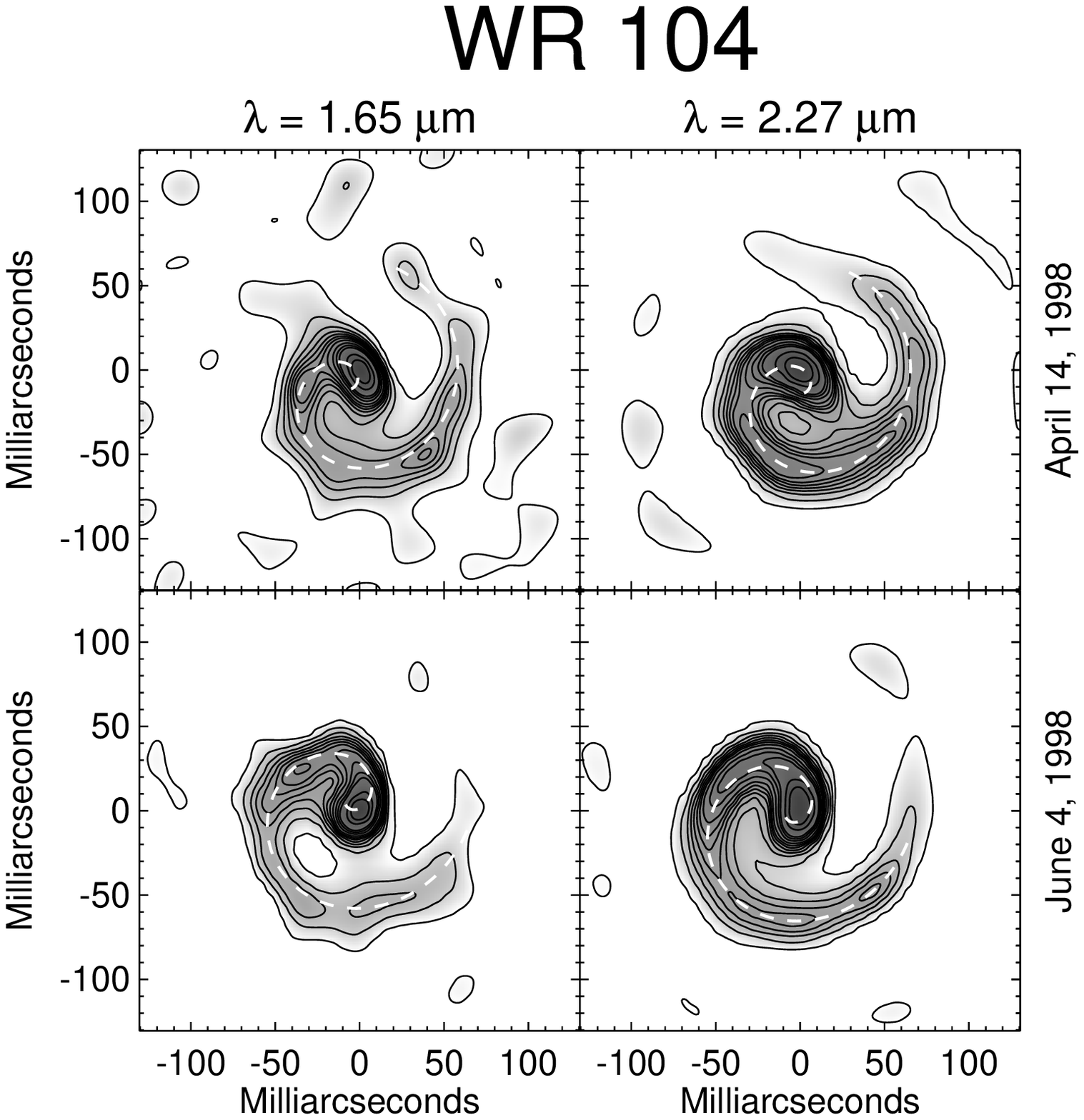}}}}
\caption{
Maps of WR\,104.
Maximum-Entropy image reconstructions of WR\,104 at 1.65 (left)
and 2.27\,$\mu$m (right) taken over two separate epochs: April (upper)
and June (lower) 1998 (JD~2450918 \& JD~2450969 respectively).
Contour levels are 0.5, 2, 4, 6, 8, 10, 12, 15, 20, 25, 30, 50 \& 70\%
of the peak.
Overplotted on each map is the best-fit Archimedian spiral model
(dashed line).
Observations utilized an annulus shaped pupil mask to form the interference
pattern, which was recorded on a fast-readout (130\,ms) infrared array
and subsequently processed to extract Fourier amplitudes and closure
phases.
Bispectral information constituting about 700 baselines
and 7000 closing triangles enabled high-quality images to be produced
from an algorithm based on the maximum entropy method \cite{GS_IEEE,Siv_84}.
Although these images are at the heart of our
discussion, it is important to note that clear and systematic signals
betraying the presence of the final image morphology are directly
visible in the calibrated Fourier data themselves.
}

\end{center}
\end{figure}

Observations of WR\,104 were made with the Keck~I telescope on 
14~April and 4~June 1998, and employed the technique of 
aperture masking interferometry in order to recover information 
out to the diffraction limit of the 10\,m Keck aperture 
\cite{HB_JOSA92,Tut_98}.
We present, in Figure~1, reconstructed images taken at
1.65 \& 2.27\,$\mu$m ($\Delta\lambda = 0.33~\&~0.16 $\,$\mu$m
respectively) for both observing epochs.
As infrared emission from the hot circumstellar dust dominates the
infrared region of the spectrum, we may interpret the highly 
asymmetric curved plumes evident in the maps as tracing the
distribution of this material.
Previous high-resolution efforts have been restricted to 
partially-resolved one-dimensional visibility curves 
interpreted in the context of spherically symmetric
outflow models \cite{Allen_MN81,Dyck_ApJ84}.
As a comparison with these earlier results, we have fitted a 
uniform-disk model to our visibilities, azimuthally averaged and
cropped to the resolutions then obtained, finding perfect agreement 
with the 130\,mas diameter disk reported in 1981 
\cite{Allen_MN81}. 
This similarity over a timescale of decades is in accord with 
the inclusion of WR\,104 in the small handful of `persistent' dust 
producing WR's \cite{Will_APSS97}.
Additional interferometric observations at 3.08\,$\mu$m 
($\Delta\lambda=0.1$\,$\mu$m) show no evidence of the marked enlargement 
towards longer wavelengths reported by Dyck et. al. \cite{Dyck_ApJ84}.
However, as is apparent from Figure~1, the images do not show even remote 
similarity to a uniform disk, and we hereafter abandon further 
consideration of circularly symmetric models.

The maps of Figure~1 consist of two components; a bright central core 
which appears elongated, and a curved tail which seems to emerge 
from one end of the elongation.
This spiral structure dominates the morphology at both colors, and maps
taken in April and June 1998 show a high degree of similarity
with the striking exception of a clear rotation of the image.
The hypothesis of dust formation mediated by the orbital motion of
a companion star and subsequently swept outwards by the stellar wind
unifies the spiral structure and the $-83^{\circ}$ rotation apparent 
between our two epochs into a simple, elegant geometry.
A schematic of our model is shown in Figure~2, showing the WR+OB binary, 
the dust formation zone associated with the collision front between the stellar 
winds \cite{Usv_MN91}, and the resultant curved outflow plume as this dust 
`nursery' is carried with the orbital motion.
Although the idea of a binary nature for WR\,104 is not new, it
is only very recently that the presence of an OB companion was
confirmed from detection of hydrogen Balmer absorption features
and optical emission-line dilution \cite{WVH_96,Crth_MN97}.

\begin{figure}
\begin{center}
\centerline{\epsfxsize=4 in{\epsfbox{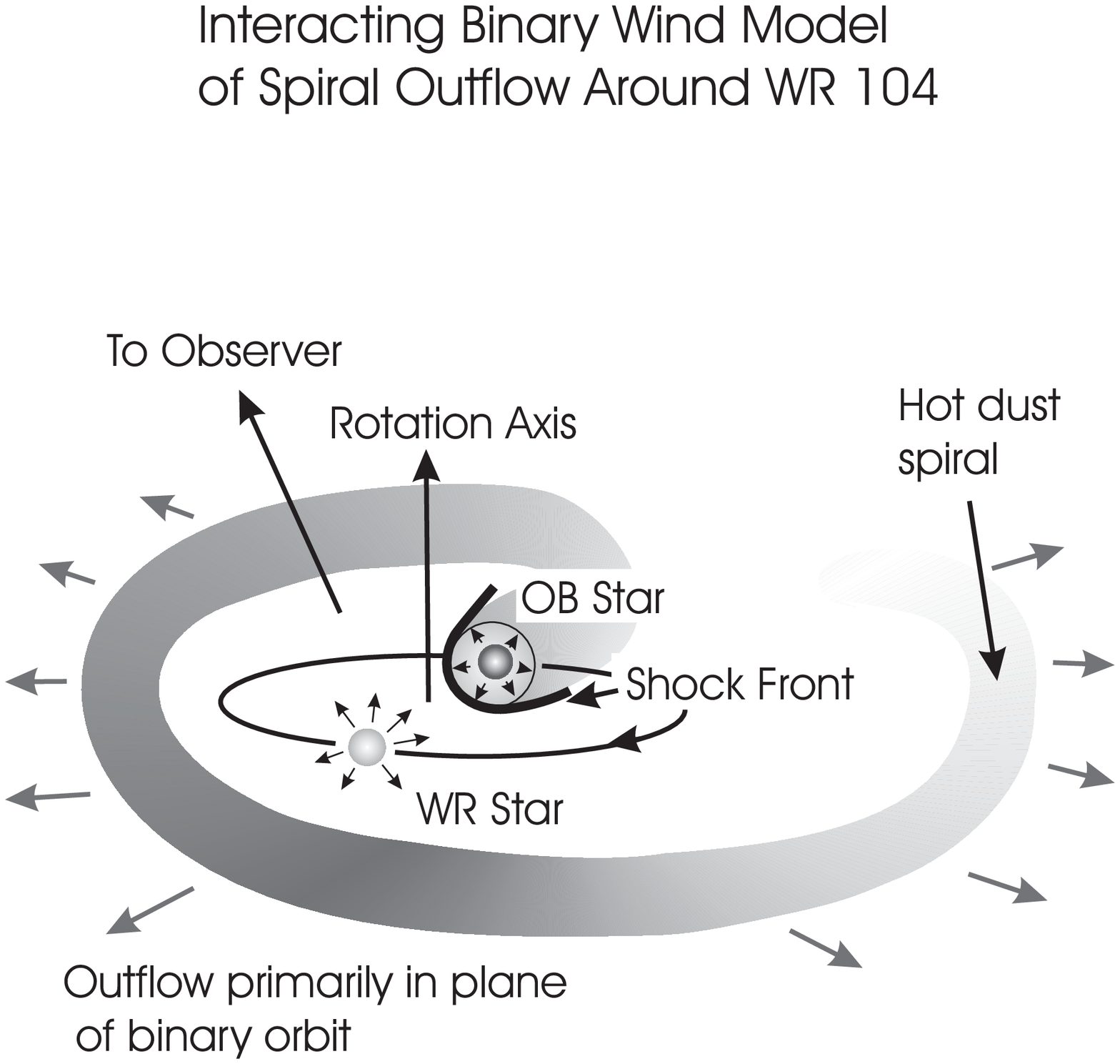}}}
\caption{
Schematic diagram of the WR\,104 binary system.
The illustration shows the WR star, the OB companion, wind-wind collision
front, and the resultant dust outflow plume (not to scale).
The spiral shape is a consequence of material being swept radially outwards
by the WR wind from a rotating dust nucleation zone associated with the
shock front where the stellar winds collide.
}
\end{center}
\end{figure}

We have overplotted, also in Figure~1, the results of fitting a simple
geometrical model consisting of an Archimedian spiral where the free
parameters are the winding rate and the viewing angle to the observer.
These modelling results were obtained by finding a global best fit to 
all four maps simultaneously, but allowing for a rotation of the spiral
structure about the model-derived axis between the two epochs.
Implicit in the assumption of an Archimedian spiral model is the 
hypothesis that the material in the spiral is moving out at a uniform
velocity, and that new material feeding into the flow insertion point 
does so at a uniform angular velocity.
Although such a model contains the fewest free parameters and yet gives
excellent fits to our data, it is important to note that a more complex model
may be required if the plume is in a zone where it is 
being accelerated, or if the orbit of the companion presumed
to be mediating the flow is eccentric (e.g. \cite{Will_APSS97}).

The physical geometry of the system, as derived directly from our
model, is a spiral plume rotating with a period of $220 \pm 30$\,days 
viewed at an angle of $20 \pm 5^{\circ}$ from the pole and with an
outflow velocity of $111 \pm 17$\,mas\,yr$^{-1}$ in the plane of the orbit.
If we identify this rotation as the orbital period of a binary stellar
system, then assuming a combined mass in the range of $20-50M_\odot$ 
\cite{MNM_ApJ90} results in a separation of $1.9-2.6$\,AU.
As this corresponds to a separation of only $\sim 1$\,mas on the sky, our images
lack the resolution to show such detail directly, and furthermore the
infrared flux is so dominated by thermal emission from the warm dust 
\cite{Zub_MN98} that it is unlikely that we have detected the central 
stars in our maps at all.
It is interesting to compare our binary parameters with those of the famous `episodic' 
dust producer WR\,140 which is known to undergo dramatic bouts of dust creation 
co-incident with the passage of a companion star through periastron in a highly 
elliptical orbit \cite{Will_APSS97}.
At periastron, the separation between the stars is $\sim 2.5$\,AU, raising
the possibility that the physical conditions favoring copious dust 
formation fall within a confined range of companion distances
for WC+OB binaries.

We may make use of the 1220\,km\,s$^{-1}$ wind outflow velocity \cite{HS_AA92} 
combined with our proper motion to derive an independent estimate of 
$2.3 \pm 0.7$\,kpc as the distance to WR\,104, where the dominant error arises 
from a $\sim$25\% uncertainty in outflow velocities found by comparing the 
results of various line-profile studies \cite{RN_ACT95} (velocities as high as 
1600\,km\,s$^{-1}$ have been reported for this star \cite{TCM_ApJ86}).
Our distance is somewhat further than earlier estimates of 1.6\,kpc 
derived from a possible association with Sgr~OB1 \cite{LuSten_AAS84}, however
the discrepancy is within the estimated errors.
Alternatively if the closer distance is preferred, then our measurements
imply an outflow velocity of 845\,km\,s$^{-1}$ for the dust component.
Our geometrical solution solves for the projected viewing
angle of the observer, and thus we avoid the usual $sin(i)$ uncertainty.
We note that although isolated dust grains should be momentum-coupled 
to the flow \cite{WHT_AA87}, the outflow velocities in the wake of
the passage of the OB~stellar companion could be significantly perturbed
and thus the behavior of the plume may not act as a good tracer of the
bulk motion of the stellar wind.
With additional observations covering an entire orbit, we will be able
to greatly refine our estimates of the physical geometry of this system.

It is apparent from Figure~1 that the outflowing material presents a relatively 
smooth, spatially confined stream without strong clumping out to a radius
of some $\sim65$\,mas ($\sim150$\,AU) from the star, by which time the outflow has
rotated through about $360^\circ$.
We believe that the finding of a single complete turn in the spiral
arm is not coincidental, as we detect only dust heated by radiation
from the central stars, and therefore lying along a direct line 
of sight.
Material in the second and further coils of the outflow will, of course,
be eclipsed by newer material closer in, and will therefore cool rapidly
resulting in the relatively sharp cutoff we see.
Although there is some evidence for brightness variations along the
arm at a level of a few percent of the peak, especially apparent in the maps taken 
at 1.65\,$\mu$m, the overall behavior points to a continuous
and smooth dust creation process, in accord with the classification
of WR\,104 as a `persistent' dust producer with a constant IR flux
\cite{ISO_AA}.
Again it is interesting to compare this behavior with that of WR\,140
whose elliptical orbit results in episodic dust production.
The contrasting characteristics of WR\,104 argue against a high degree
of orbital eccentricity, giving some justification to our choice of
the Archimedian spiral model in this case.

For WR\,104, our observations confine the IR excess emission from the dust 
to lie in a narrow, spatially confined outflow which rotates synchronously 
with a period of 220 days -- a plausible period for a wind-interacting
binary system.
No spherically symmetric or diffuse component to the dust nebula was detected
to within a few percent of the peak flux.
We are therefore able to reject dust-formation models resulting in spherical 
or disk shaped outflows such as the clumpy spherical outflows of \cite{Veen_AA98} 
or equatorial density enhancements \cite{CIB_ApJ96} in favor of the binary 
wind-wind model.

The viewing angle to the observer of $20 \pm 5^{\circ}$  is well constrained
by these measurements.
This finding of an almost face-on system contradicts previous attribution of
high circumstellar extinction \cite{Coh_AA75} and spectral variability 
\cite{Crth_MN97} to an edge-on viewing angle.
Some of these observations may be explained as WR~104 is thought to lie behind
a heavily obscuring cloud \cite{LuSten_AAS84} with further extinction possibly
arising from material created in past mass-loss events of the progenitor star.

As the dust comprises only a very small fraction of the total mass loss, it
therefore acts as a visible tracer in the outflow enabling the fascinating
possibility of dynamical studies of the wind itself.
Detailed numerical modelling is needed to determine if the high degree
of initial collimation and subsequent confinement of the dust plume can be
explained with simple models of the wind-wind interaction
\cite{Usv_MN91}, or whether more detailed three-dimensional calculations such 
as those of Walder \cite{Wal_IAU163}, are required. 
Spectral studies of the plume, beyond the scope of this letter, should reveal
the thermal and chemical evolution of the dust as it is swept outwards into the
interstellar medium, and also yield information on processes underlying the
binary-mediated dust creation mechanism.
With a handful of dusty WR systems open to study with this novel method for 
the detection of binary stars, wider questions of dust formation in this 
class of objects can now be addressed.

{\bf Acknowledgements} Data herein were obtained at the W.M. Keck 
Observatory, made possible by the generous support of the W.M. 
Keck Foundatation, and operated as a scientific partnership among the
California Institute of Technology, the University of California and NASA.
This work was supported through grants from the National Science Foundation.
The authors would like to thank Devinder Sivia for the maximum-entropy
mapping program ``VLBMEM'' and David Hale for sparking our interest in
Wolf-Rayet stars.

All correspondence should be addressed to Peter Tuthill 
(e-mail:gekko@ssl.berkeley.edu)

\end{document}